\documentclass[aps,prl,reprint,superscriptaddress]{revtex4-1} 
\usepackage{graphicx}
\usepackage{amsmath,amssymb}
\usepackage{url}

\newcommand{\tss}[1]{$_{\mathrm{#1}}$}

\providecommand{\abs}[1]{\lvert#1\rvert}

\begin{document}
\title{Delocalised oxygen as the origin of two-level defects in Josephson junctions}

\author{Timothy C. DuBois}
\affiliation{Chemical and Quantum Physics, School of Applied Sciences, RMIT University, Melbourne, 3001, Australia}

\author{Manolo C. Per}
\affiliation{Chemical and Quantum Physics, School of Applied Sciences, RMIT University, Melbourne, 3001, Australia}
\affiliation{Virtual Nanoscience Laboratory, CSIRO Materials Science and Engineering, Parkville, VIC 3052, Australia}

\author{Salvy P. Russo}
\affiliation{Chemical and Quantum Physics, School of Applied Sciences, RMIT University, Melbourne, 3001, Australia}

\author{Jared H. Cole}
\affiliation{Chemical and Quantum Physics, School of Applied Sciences, RMIT University, Melbourne, 3001, Australia}

\date{\today}

\begin{abstract}

One of the key problems facing superconducting qubits and other Josephson junction devices is the decohering effects of bi-stable material defects. Although a variety of phenomenological models exist, the true microscopic origin of these defects remains elusive. For the first time we show that these defects may arise from delocalisation of the atomic position of the oxygen in the oxide forming the Josephson junction barrier. Using a microscopic model, we compute experimentally observable parameters for phase qubits. Such defects are charge neutral but have non-zero response to both applied electric field and strain. This may explain the observed long coherence time of two-level defects in the presence of charge noise, while still coupling to the junction electric field and substrate phonons.

\end{abstract}

\maketitle

Decoherence is currently a major limitation for superconducting qubits and Josephson junction based quantum devices in general.  An important source of decoherence stems from environmental two-level systems~\cite{Dutta81, Shnirman05}. Recent experiments have even probed these defects directly and shown that they are stable, controllable and have relatively long decoherence times themselves~\cite{Simmonds04, Neeley08, Shalibo10, Lupascu09, Lisenfeld10}. Little is known about the true microscopic nature of these defects, although many phenomenological theories exist~\cite{Martinis05, deSousa09, Sendelbach08, Faoro07, Ku05}. We take a novel approach to the problem: starting from atom positions and species, motivated by \emph{ab initio} and molecular mechanics methods. Using this approach we compute experimentally observed parameters such as resonant frequency, defect-qubit coupling and response to strain, and find excellent agreement with experiments. We show that the quantum property of delocalisation of the oxygen \emph{atomic position} in aluminium oxide naturally results in a model for two-level defects without the need for additional impurities.

The existence of bistable defects in glasses and amorphous solids in general is well known~\cite{Anderson72}. Amorphous insulating barriers (either in the form of Josephson junctions (JJ) or simply a native oxide) form an integral part of superconducting circuits, so it comes as no surprise that two-level systems (TLSs) are often considered to be an important source of noise in these circuits~\cite{Dutta81, Shnirman05, Martinis05}.  The recent development of controllable qubit circuits (charge, flux or phase) has provided the opportunity to study so-called `strongly coupled defects'~\cite{Neeley08, Lupascu09, Lisenfeld10}. These defects have comparable resonance frequencies to the qubit circuit and coupling strengths and decoherence times large enough to allow coherent oscillations between qubit and TLS.  Probing individual defects has promoted their bistable nature from hypothesis to observable fact as well as providing clues to their microscopic origin.

There exists an embarrassment of riches in terms of theoretical models for such defects. Various phenomenological models exist, including charge dipoles~\cite{Martinis05}, Andreev bound states~\cite{deSousa09}, magnetic dipoles~\cite{Sendelbach08}, Kondo impurities~\cite{Faoro07} and TLS state dependence of the JJ transparency~\cite{Ku05}.  Although detailed fitting of experimental data can place limits on these models~\cite{Cole10}, they all have enough scope within their free parameters to explain the observed behaviour - rendering them presently indistinguishable.

To make concrete predictions, a detailed microscopic model of these defects is required. In this paper we consider the origin of defects to be within the amorphous oxide layer itself~\cite{Lacquaniti12}, rather than assuming defects stemming from surface states~\cite{Choi09} or the accidental inclusion of an alien species. A pertinent example defect is the oxygen interstitial in crystalline silicon. For an O defect in c-Si, the harmonic approximation for atomic positions cannot be applied due to the rotational symmetry of the defect as oxygen delocalises around the Si-Si bond axis~\cite{Artacho95}. This forms an anharmonic system with a quasi-degenerate ground state, even in a ``perfect'' crystal. As many different spatial configurations can exist in the AlO$_{x}$ amorphous junction, it is our premise that positional anharmonicity arises within voids in this layer.  This yields TLSs with unique properties based solely on atomic positions and rotation in relation to the external electric field. Starting from this ansatz, we compute parameters which have been measured directly in experiments on TLSs, including: TLS resonant frequency, qubit-TLS coupling and TLS energy/strain dependence.

\begin{figure}
\centering
 \includegraphics[width=\columnwidth]{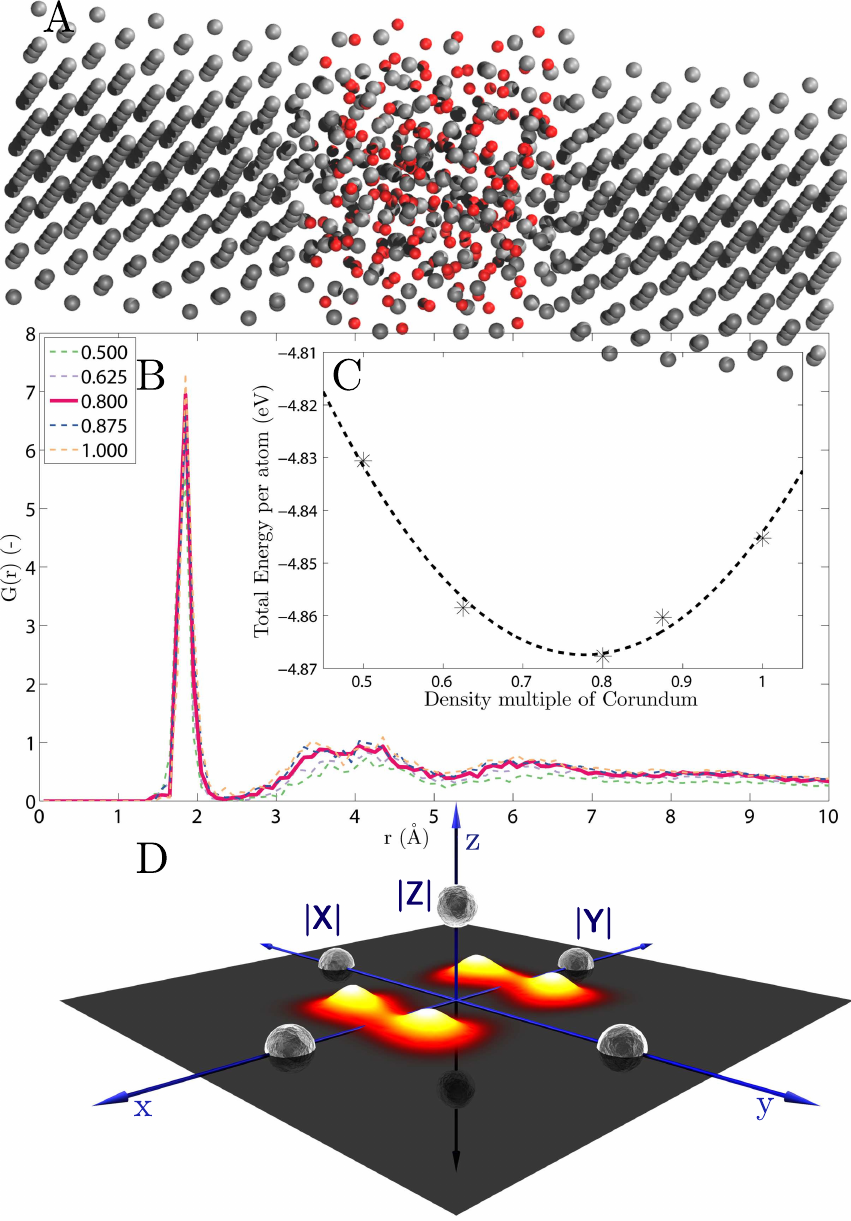}
 \caption{\label{fig:1}a) Depicts a JJ with two aluminium slabs surrounding an amorphous AlO$_{1.25}$ barrier (aluminium: gray, oxygen: red). b) The projected radial distribution function $G(r)$ using oxygen as a reference. c) Shows the total energy per atom of this structure as a function of oxide density. Fluctuations in the fine structure of b and c are due to finite box restrictions of the model. d) An illustration of the 2D oxygen delocalisation model. Aluminium atoms in gray, with the delocalised oxygen atom probability density shown for an example ground state distribution.}
\end{figure}

To begin our investigation, a JJ was modelled using molecular mechanics and Density Functional Theory (DFT). This approach has previously been used to model defects and dopants in aluminium oxide~\cite{Kim09,Kim09a}.  A representation of our system after amorphisation of the AlO\tss{1.25} barrier layer is displayed in Fig.\ \ref{fig:1}a. Experimental O/Al ratios have been shown to be highly dependent on fabrication processes~\cite{Park02, Tan05}, therefore a representative stoichiometry of AlO\tss{1.25} was chosen for simplicity. Various values of the oxide density were calculated and $3.2 \; \rm{g/cm}^3$ was found to minimise the total energy of the system (Fig.\ \ref{fig:1}c), which agrees with experimental values~\cite{Barbour98}. This density is $0.8$ times that of a common crystalline form of Al\tss{2}O\tss{3} (Corundum). Using these simulations, we compute the projected radial distribution function $G(r)$ (Fig.\ \ref{fig:1}b) of the resulting atomic positions. Both the Corundum peak ($\sim 1.97 \; \rm{\AA}$) and a broad distribution ($> 3 \; \rm{\AA}$) corresponding to the amorphous AlO$_{x}$ layer are visible. Details of the precise parameters used in the DFT calculations can be found in the supplementary material.

The energy scale for JJ defects observed in experiments is $\lesssim 40 \; \rm{\mu eV}$~\cite{Neeley08, Lupascu09, Lisenfeld10}. This energy splitting, while large for qubit experiments, is very small when compared to typical electronic structure calculations, i.e.\ the ground and first excited state of our defect form a quasi-degenerate ground state on the scale of crystal defect energies, which puts them below the precision limits of DFT. Using the $G(r)$ data obtained from DFT as a starting point, we develop an effective single-body model. This allows higher precision calculations as a function of atom locations, using empirical potentials for the interactions between an oxygen and nearest neighbour aluminium atoms. We initially consider a cubic lattice of six aluminium atoms with an oxygen atom delocalised at its center as our prototype defect (Fig.\ \ref{fig:1}d displays a representation of the case for delocalisation in 2D). As the experimental results point to bistable defects, we assume that the observed behaviour does not rely on a spherically symmetric potential in all three spatial dimensions, which would lead to triple degeneracy. We therefore concentrate on two-dimensional delocalisation. Using the empirical Streitz-Mintmire potential~\cite{Streitz94} we derive an effective single particle Hamiltonian
\begin{equation}
    H = -\frac{\hbar^2}{2m_{oxy}}\nabla^2+V(\mathbf{r}),
    \label{eq:OHam}
\end{equation}
where $m_{oxy}$ is the mass of an oxygen atom and $V(\mathbf{r})$ is the potential due to the six aluminium atoms, generated by the Streitz-Mintmire formalism. The resulting single-body time-independent Schr\"odinger equation is then solved on a finite grid (see supplementary material). 

For our effective model, the atomic positions $\abs{X}$, $\abs{Y}$ and $\abs{Z}$ labeled in Fig.\ \ref{fig:1}d represent aluminium atom pairs (e.g.\ $-X$,$+X$) lying on the cardinal axes and displaced equidistantly from the origin in each direction.
%

\begin{figure*}
\centering
   \includegraphics[width=\textwidth]{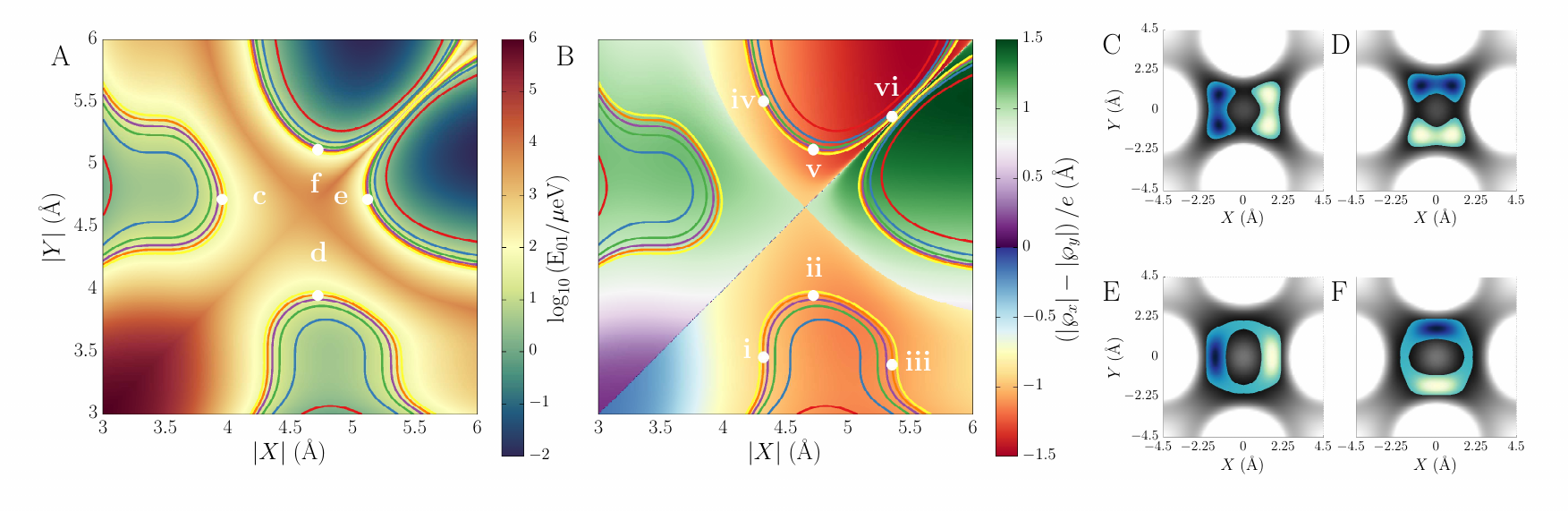}
 \caption{\label{fig:2}a) Map of the $E_{01}$ energy splittings of the delocalised oxygen 2D model. The $\abs{X}$ and $\abs{Y}$ axes represent aluminium pair positions with $\abs{Z} = 2.5788 \; \rm{\AA}$. b) The difference between the absolute dipole moment (in $x$- and $y$-directions) over the same range. We see either $\abs{\wp_x}$ (red) or $\abs{\wp_y}$ (blue) dominated behaviour in all regions except $\abs{X},\abs{Y} \lesssim 3.5 \; \rm{\AA}$ (where the oxygen is tightly confined). c) through f) show the first excited state wavefunction of the oxygen atom and the acting potential of four configurations indicated on a). For comparison with existing qubit experiments we plot contour lines corresponding $E_{01}/h = 0.5 - 10$ GHz (red to yellow) overlayed on a) and b). The same resonant frequencies are discussed in Fig.\ \ref{fig:3}, hence c-f all represent configurations of $E_{01}/h = 8$ GHz; whereas points i-vi are the locations selected for the strain study in Fig.\ \ref{fig:4}.}
\end{figure*}

In two dimensions, the potential landscape of interest approximates the venerable ``Mexican hat'' potential as this is the rotationally symmetric generalization of the double-well potential in 1D. This potential has a unique, spherically symmetric ground state and a doubly degenerate first excited state, which we see for $\abs{X}=\abs{Y}>\abs{Z}$. Small deformations due to $\abs{X}\neq\abs{Y}$ or translations of $\abs{X}$ or $\abs{Y}$ off axis quickly result in a quasi-degenerate ground state of the form seen experimentally. The resulting splitting, $E_{01}$, is plotted in Fig.\ \ref{fig:2}a for a range of $\abs{X}$ and $\abs{Y}$ values, with a fixed value of $\abs{Z}=2.5788\;\rm{\AA}$. Fig.\ \ref{fig:2}b shows the absolute dipole moment response over the same phase space. Subfigures \ref{fig:2}c-f display four positions of equal energy on Fig.\ \ref{fig:2}a, where the first excited-state wavefunction of the oxygen atom is plotted with its effective potential.

The phase space in figures \ref{fig:2}a and \ref{fig:2}b are split into four domains, \ref{fig:2}a via dispersive peaks with their maxima located at the bifurcation lines visible on \ref{fig:2}b. The properties of these domains can be explained through the interplay of potential configuration and dipole alignment.  In two-dimensions these potentials can be described as a sum of two 1D potentials: one in the $x$ direction and the other along $y$. The first case is a sum of two double wells (tetra-well) and the second, a sum of a double and harmonic well (hemi-tetra-well). Points c and d in Fig.\ \ref{fig:2} are examples of the former while e and f represent the latter.  Similarly, each domain has a dipole element which is orientated in either $x$ or $y$, as one observes from the direction of the nodal line in each subfigure. These computed dipole moments correspond well to observed values, assuming $\mathcal{O}\left(\rm{nm}\right)$ junction widths~\cite{Martinis05, Cole10}.

To compare our TLS model directly to experiments, we assume that our JJ lies within a phase qubit, although the model applies equally for any device comprised of amorphous junctions. The measurable signal of a TLS in a phase qubit is the resonance of the TLS and qubit splitting energy, $E_{01}$, with the qubit-TLS coupling, $S_{max}$. For the phase qubit~\cite{Martinis05}, $S_{max}$ is a function of $E_{01}$ and $\wp$~\cite{Kofman07}, the effective dipole moment due to an electric field applied in the direction of delocalisation,

\begin{equation}
    S_{max}=2\frac{\wp}{w}\sqrt{\frac{e^2}{2C}E_{01}}
\label{eq:smax}
\end{equation}
(see supplementary material). Throughout this discussion we assume a junction width $w = 2$ nm and capacitance $C = 850$ fF.

\begin{figure}
\centering
 \includegraphics[width=.8\columnwidth]{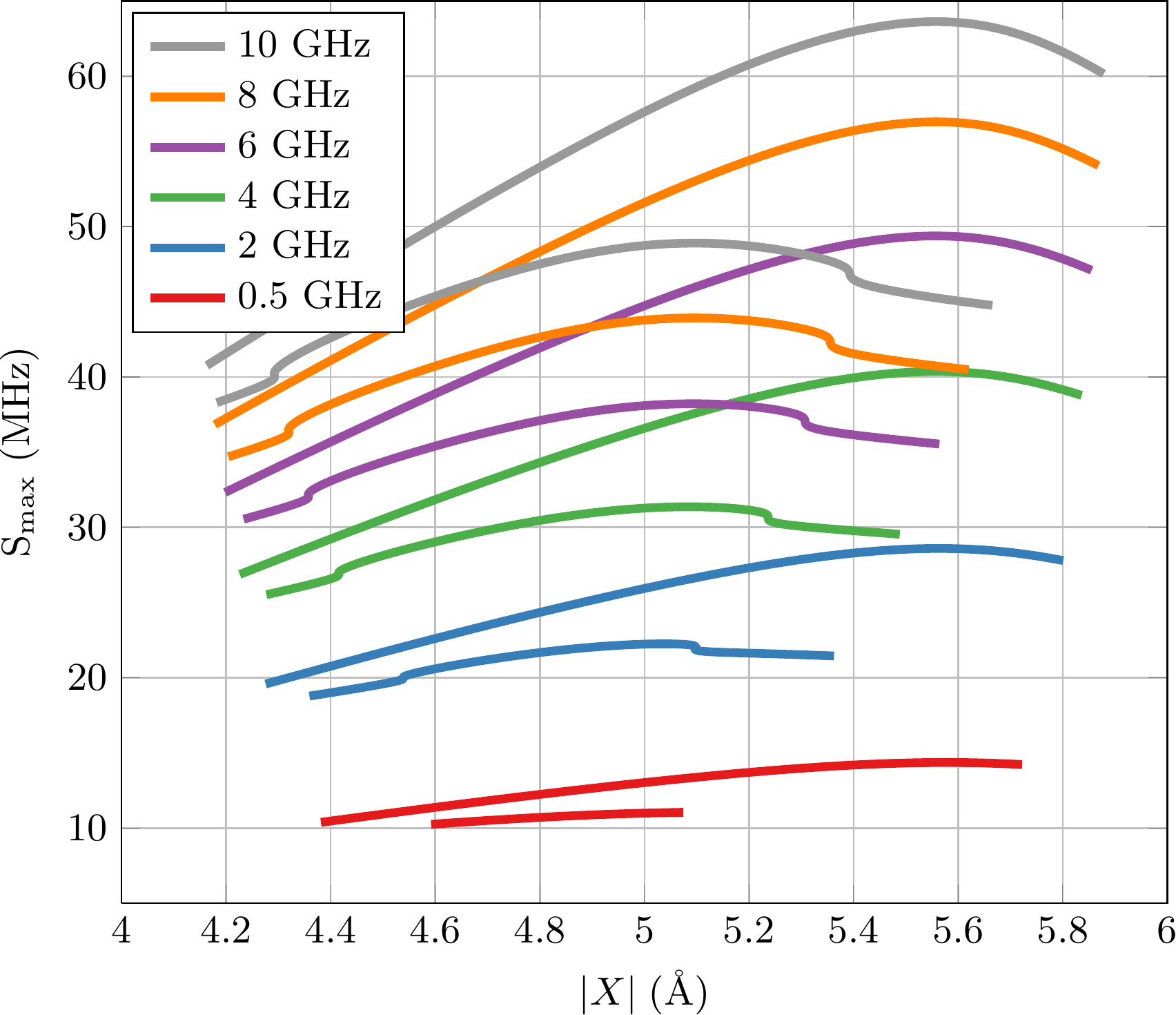}
\caption{\label{fig:3}Coupling strength to a fictitious phase-qubit $S_{max}$ as a function of $\abs{X}$ in the domains where $\abs{\wp_x}$ is dominant (see Eq.\ \ref{eq:smax}) for a set of constant $E_{01}$ splitting frequencies. For comparison with experimental results, $E_{01}$ and $S_{max}$ are expressed in frequency units.}
\end{figure}

In Fig.\ \ref{fig:3} we plot contour lines representing constant values of $E_{01}$ which correspond to the purview of experimentally observed qubit resonant frequencies. This region of parameter space corresponds well with the calculated $G(r)$ (see Fig.\ \ref{fig:1}c) in that spacings of $\abs{X}=4-6 \; \rm{\AA}$ are possible, albeit uncommon. This graph shows the $\abs{X}$ response in the domains where $\abs{\wp_x}$ is dominant, although it is clear to see that the $\abs{Y}$ domains are symmetric from Fig.\ \ref{fig:2}. The $S_{max}$ (Eq.\ \ref{eq:smax}) response to these frequencies is plotted as a function of $\abs{X}$, in which we see maximum coupling strengths which correspond well with experimental observations~\cite{Lupascu09, Shalibo10, Cole10}. Two contours are visible for each $E_{01}$ splitting, which is due to the bifurcation in Fig.\ \ref{fig:2} as the values of $\abs{X}$ and $\abs{Y}$ flip in phase space - indicating a change from the tetra- to the hemi-tetra region.

A key observation of the TLS-qubit experiments is the unusually long coherence times of strongly coupled defects~\cite{Neeley08, Lisenfeld10a}. As our model assumes a charge-neutral defect, coherence time is linked to the dipole element (for charge noise) and the strain response (for phonons). The strain response has recently been observed directly through mechanical deformation of a phase-qubit~\cite{Grabovskij12}.

\begin{figure}
\centering
 \includegraphics[width=\columnwidth]{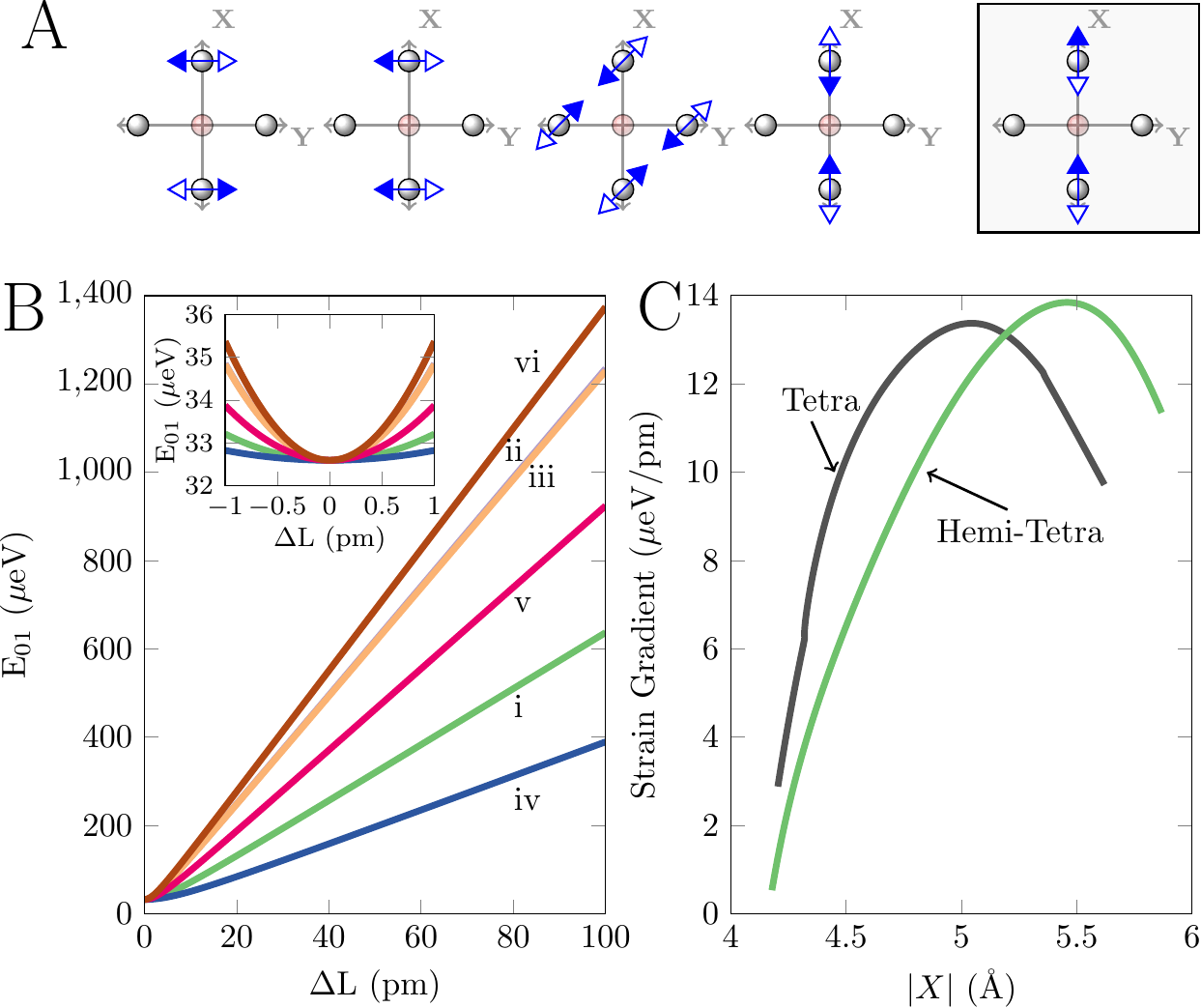}
\caption{\label{fig:4}a) depicts a number of deformations which were applied to the aluminium atoms in the $x-y$ plane (see text). Only one of the modes responded with over a few Hz of movement (highlighted), which is indicative of an optical phonon mode, generating frequency splittings of $\mathcal{O}\left(100 \; \rm{MHz}\right)$ for picometer deformations. A deformation range of $100$ pm was applied to the points i-vi from Fig.\ \ref{fig:2} yielding responses that are both hyperbolic and symmetric (b). Over the range $20 - 100$ pm the response is linear with a strain gradient, shown in c) for both the tetra- and hemi-tetra domains in the $\wp_x$-direction.}
\end{figure}

We introduce a series of deformations in our 2D model to measure the variation in $E_{01}$, which are depicted in Fig.\ \ref{fig:4}a. All deformations were tested in each of the four regions of Fig.\ \ref{fig:2}a, not only in the $x$-direction as shown, but also in $y$. Of the tested deformations we find the response of one (the optical phonon mode, highlighted in Fig.\ \ref{fig:4}a) to be $10^5$ times stronger than the others. Such a deformation corresponds to a translation of both aluminium atoms in the same direction and relative to the oxygen, along the axis of the dominant dipole. This suggests an explanation for the long TLS coherence times, as a delocalised oxygen is only sensitive to a small subset of available phonon modes (as well as coupling to charge noise only through its electric dipole). Fig.\ \ref{fig:4}b shows this response for the optical mode (at several points of interest labeled in Fig.\ \ref{fig:2}b), displaying a characteristic hyperbolic response which is typical of a two-level system. This compares well with the observed strain response in Ref.\ \onlinecite{Grabovskij12}. Finally, Fig.\ \ref{fig:4}c shows the linear strain gradient plotted along the $E_{01}/h = 8$~GHz contour for the tetra- and hemi-tetra- regions in the $\abs{\wp_x} \neq0, \abs{\wp_y}=0$ domains.

Our model allows prediction of experimentally measured properties of strongly coupled TLSs with atomic positions as the only input parameters. Using realistic atomic positions obtained via molecular mechanics and \emph{ab initio} methods, the correspondence with observed defect properties is excellent and therefore suggests that these defects can arise in AlO$_{x}$ without any alien species present. Our model also proposes that restricting the delocalisation of oxygen, for example through higher densities in the amorphous layer, results in fewer voids~\cite{Lacquaniti12} and therefore fewer two-level defects. Microscopic models of this type will guide future fabrication and design of superconducting circuits, leading to lower levels of noise and greater control over their quantum properties.

\bibliography{delocal}

\end{document}